\documentclass[preprint,aps,showpacs]{revtex4}
\usepackage [cp1251]{inputenc}
\usepackage{amsmath}
\usepackage{bm}

\usepackage{color}
\usepackage{epsfig}

\newcommand{\nix}[1]{}

\begin{document}

\title{Spin-dependent recombination and hyperfine interaction at the deep defects}
\author{E.L. Ivchenko, L.A. Bakaleinikov, V.K. Kalevich}
\affiliation{Ioffe Physical-Technical Institute, 194021 St. Petersburg, Russia}
\begin{abstract}
We present a theoretical study of optical electron-spin orientation and spin-dependent Shockley--Read--Hall
recombination taking into account the hyperfine coupling between the bound-electron spin
and the nuclear spin of a deep paramagnetic center. We show that the number of master rate equations
for the components of the electron-nuclear spin-density matrix is considerably reduced due to the restrictions
imposed by the axial symmetry of the system under consideration. The rate equations describe the Zeeman
splitting of the electron spin sublevels in the longitudinal magnetic field,
the spin relaxation of free and bound electrons, and the nuclear spin relaxation in the two defect states,
with one and two (singlet) bound electrons. The general theory is developed for an arbitrary value of the nuclear spin $I$,
the magnetic-field and excitation-power dependencies of the electron and nuclear spin polarizations are calculated
for the particular value of $I=1/2$. The role of the nuclear spin relaxation in each of the both defect states is analyzed.
The circular polarization and intensity of the edge photoluminescence as well as the dynamic nuclear spin polarization as functions of the excitation power
are shown to have bell-shaped forms.

\end{abstract}

  \maketitle

\section{Introduction}
Spin-dependent recombination (SDR) via deep paramagnetic centers
has recently attracted increased interest and proved to be an
effective tool for obtaining an abnormally high spin polarization
of free and bound electrons in nonmagnetic semiconductor alloys
GaAs$_{1-x}$N$_x$, Ga$_{1-y}$In$_y$As$_{1-x}$N$_x$ and
semiconductor quantum wells Ga(In)AsN/GaAs at room temperature
\cite{JETP1,JETP2,apl2009,current,Int1,QW2010}, see also
\cite{Condensed} and references therein. The centers occupied by
single spin-polarized electrons act as a spin filter
\cite{Weisbuch,Paget,Condensed} and block the free electrons of
the same spin polarization from escaping from the conduction band.
As a result the spin polarization of free photoelectrons generated
by circularly polarized optical excitation (as well as that of bound electrons)
can be enhanced up to $\sim$100\%. The amplification of spin polarization is accompanied
by an increase in the concentration of photoelectrons, the
intensity of band-to-band photoluminescence (PL) and the
photoconductivity, as compared to the linearly polarized
photoexcitation \cite{JETP1,apl2009,current,PRBcond}.

The hyperfine interaction between the localized electron and the
nucleus of the deep center mixes their spin states resulting in
(i) a reduction of the initial electron spin polarization and (ii)
dynamic nuclear polarization of the defect atoms \cite{DP1972}. In
the absence of an external magnetic field, the localized-electron
spin polarization can be reduced down to 1/2 and 3/8 for the
nuclear spin $I = 1/2$ and 3/2, respectively. The longitudinal
magnetic field suppresses the hyperfine coupling and restores the
electronic polarization as soon as the electron Zeeman energy
exceeds the hyperfine interaction: The expected increase in the
intensity and circular polarization of the edge PL
in the longitudinal magnetic field has been confirmed
experimentally. In addition, strong nuclear polarization effects,
due to a combination of the spin-dependent recombination and
hyperfine coupling, have been reported and discussed in Refs.
\cite{naturemat,KalevichPRB,JetpLett2012,APL2013,Buyanova2013,NatureComm,Chen2014,Toulouse2014}.
Particularly, the dynamically polarized nuclei create an effective
magnetic field (the Overhauser field) acting on the spins of
localized electrons; this field is added to the external magnetic
field and shifts the `electron polarization vs. field' curve, with
the shift changing the sign under reversal of the circular
polarization of the exciting light
\cite{KalevichPRB,JetpLett2012,NatureComm}.

The theory of spin-dependent Shockley--Read--Hall recombination derived
in Ref.~\cite{JETP2}, see for more detais \cite{Condensed},
ignores the nuclear effects. It has been successfully applied to
describe the main features of optical spin orientation of
conduction-band and deep-level electrons in GaAsN at zero and
transverse magnetic field ${\bm B} \perp z$, where the axis $z$ is
parallel to the exciting light beam and coincides with the normal
to the sample surface. The model of Ref.~\cite{JETP2} is unable to
interpret the experimental data obtained in the longitudinal
magnetic field ${\bm B} \parallel z$. The initial way out
\cite{KalevichPRB} was to assume the spin-relaxation time
$\tau_{sc}$ of bound electrons to depend on the magnetic field
$B_z$. This assumption could explain the polarization recovery
with increasing the field but faced with the pressing need to find
a mechanism of the field dependence of $\tau_{sc}$ which looked
unresolvable. Moreover, the modified model cannot provide a
reasonable interpretation of the observed shift of the
polarization-field curve changing the sign under the reversal of
circular polarization of the incident light.

The first attempt to give a theoretical description of the studied
nuclear polarization processes has been performed by Puttisong
{\it et al}., see Supplementary Methods for  Ref.~\cite{NatureComm}. In
that work, the hyperfine interaction is taken into account
approximately by introducing magnetic-field-independent flip-flop processes in the
electron-nuclear system and including an additional
phenomenological parameter, the flip-flop spin relaxation time.
This approximation obviously provides physical insight into the
role of the nuclei but its validity for a quantitative description
is not obvious. A kinetic theory of the spin-dependent
recombination incorporating the hyperfine interaction of
electronic and nuclear spins has been proposed recently by
Sandoval-Santana {\it et al}. \cite{Toulouse2014} The master equation
approach for the spin-density matrix of the electron-nuclear
system includes 144 equations which are solved numerically. The
numerical calculation reproduces the main experimental findings of
Ref.~\cite{JetpLett2012}. Nevertheless, the role of spin
relaxation of nuclei in the system under consideration still
remains open. In Ref.~\cite{Toulouse2014} the nuclear spins are
polarized only in the deep-center states with single bound
electrons. The nuclei with two bound electrons are characterized
just by their steady state average concentration $N_2$. This means
nothing more than that the formulation of Ref.~\cite{Toulouse2014}
is based on the assumption of very fast nuclear spin relaxation in
the defect state with a pair of electrons. As far as we know, at
present there are no grounds to take this assumption for granted.
In general the spin relaxation times $\tau^{(1)}_n$ and
$\tau^{(2)}_n$ for defect states with one and two bound electrons
can be of the same order and even longer than the lifetimes of
these states. In this work we develop a theory of the
spin-dependent recombination and hyperfine coupling for the
arbitrary values of $\tau^{(1)}_n$ and $\tau^{(2)}_n$. The paper
is organized as follows. In Sec.~II we introduce the
electron-nuclear spin-density matrix of the defect state with a
single bound electron and the spin-density matrix of the defect
with two bound electrons (in the singlet state) and discuss the
restrictions imposed on the nonzero components of these matrices
by the axial symmetry of the system in the longitudinal magnetic
field. In Sec.~III, we derive the rate equations for the spin
density matrices taking into account both the hyperfine coupling
for a nucleus with the angular momentum $I$ and the electron and
nuclear spin relaxation. The particular limiting cases are
analyzed in Secs.~III A, B and C. The simplifications in the
case of a nucleus with $I=1/2$ are considered in Sec.~IV. The
results of numerical calculation and their discussion are
presented in Sec.~V. Section VI contains the concluding remarks.

\section{Electron-nuclear spin-density matrix}
We use the basic states $|s,m \rangle$ of the electron-nuclear system, where $s=\pm 1/2$ and $m$ ($- I \leq m \leq I$) are the bound-electron and nuclear spin projections upon the fixed axis $z$, hereafter the normal to the sample surface, and $I$ is the angular momentum of a nucleus. In the first, general, part of the paper we will take $I$ to be arbitrary and then shift to the particular case of $I=1/2$ which allows simplification of the kinetic equations for the densities and spin polarizations of the free and bound electrons. For the deep defect responsible for the spin-dependent recombination in GaAs$_{1-x}$N$_x$, the momentum $I$ is 3/2. A detailed analysis for this value of $I$ will be performed elsewhere.

In addition to $|s,m \rangle$, we also use the notation $|s, M - s \rangle$ for the state with the electron spin $s$ and the total component of the angular momentum $M = s + m$. In the following we take into account the hyperfine interaction of the electron and nuclear spins given by the Fermi contact Hamiltonian
\[
{\cal H}_{\rm hf} = A~{\bm s} \cdot {\bm I}\:,
\]
where $s_{\alpha}$ and $I_{\alpha}$ ($\alpha = x,y,z$) are the electron and nuclear spin operators. Moreover, we consider  the normal incidence of the polarized exciting light in the external magnetic field ${\bm B} \parallel z$ (Faraday geometry), take into account the Zeeman Hamiltonian ${\cal H}_{\bm B} = g \mu_B B_z s_z$ for the bound electrons and neglect the interaction between the magnetic field and the magnetic moments of the nuclei or conduction-band electrons. Here the bound-electron  Land\'e factor $g \approx 2$ and $\mu_B$ is the Bohr magneton.

The occupation of the defect with one bound electron is described by a $2(2I + 1)\times 2(2I + 1)$ spin-density matrix $\rho_{sm,s'm'}$. In the Faraday geometry, the components with unequal total angular-momentum components $M = s+m$ and $M' = s' + m'$ vanish. Therefore, it is enough to consider the components
\begin{equation} \label{rhoM}
\rho_{s,M - s;s',M - s'} \equiv \rho_{ss'}^{(M)}\:,
\end{equation}
which are normalized on the density of single-electron defects
\[
\sum\limits_{s,m} \rho_{sm,s m} = \sum\limits_{s M} \rho_{ss}^{(M)} = N_1\:.
\]
The matrix $\rho_{ss'}^{(M)}$ with $M=I + 1/2$ or $M=-(I + 1/2)$ contains only one non-zero component
and can be presented as
\[
\rho_{ss'}^{(I + \frac12)} = \delta_{ss'} \delta_{s, \frac12} \rho_{\frac12, I;\frac12, I} \hspace{2 mm} \mbox{and} \hspace{2 mm} \rho_{ss'}^{(-I - \frac12)} = \delta_{ss'} \delta_{s, -\frac12} \rho_{-\frac12,-I; - \frac12, - I}\:.
\]
It is worth to note that the electron spin-density matrix (2$\times$2 matrix)
\begin{equation} \label{rhoe}
\rho^e_{ss'} = \sum\limits_M \rho_{ss'}^{(M)}
\end{equation}
is diagonal whereas the matrices $\rho_{ss'}^{(M)}$ with $|M| < I + 1/2$ contain off-diagonal components. In the geometry under consideration, the spin-density matrix of the defect singlet with two bound electrons is diagonal, its $2I + 1$ diagonal components $N_{2,m}$ are normalized on the density of double-electron defects, $N_2$. The sum of $N_1$ and $N_2$ gives the density of deep defects, $N_c$.

Thus, for a nucleus with $I=3/2$, instead of 144 equations declared in Ref. \cite{Toulouse2014} there are only 21 nonzero quantities to be found: 2 components $\rho_{1/2,1/2}^{(2)}$ and $\rho_{-1/2,-1/2}^{(-2)}$, 12 components $\rho_{s,s'}^{(M)}$ with $M = 0, \pm 1$ and $s, s' = \pm 1/2$, four components $N_{2,m}$ with $m=\pm 3/2, \pm 1/2$, the densities of electrons $n_{\pm 1/2} \equiv n_{\pm}$ in the conduction band with the spin $\pm 1/2$ and the unpolarized free-hole density $p$.
\section{Kinetic equations for the spin-density matrix}
The two kinetic equations
\begin{eqnarray} \label{twofirst}
&&2c_n N_- n_+ + \frac{n_+ - n_-}{2 \tau_s}= G_+\:, \label{1a}\\
&&2c_n N_+ n_- + \frac{n_- - n_+}{2 \tau_s}= G_- \label{1b}
\end{eqnarray}
have the same form as those in the model of Ref.~\cite{Condensed} where the hyperfine coupling was ignored.
Here $N_+ = \rho^e_{1/2, 1/2}$ and $N_- = \rho^e_{-1/2, -1/2}$ are the densities of single-electron defects with the electron spin $\pm 1/2$, their sum $N_+ + N_-$ being $N_1$,
$G_+$ and $G_-$  are the generation rates of the spin-up and spin-down photoelectrons, and $c_n$ is the proportionality constant in the electron trapping rate by deep centers.
We remind that, due to the relations
\begin{eqnarray}
&&N_+ + N_- + N_2 \equiv N_1 + N_2 = N_c\:, \label{densities1}\\
&&p = n + N_2\:,\: n = n_+ + n_-\:, \label{densities2}
\end{eqnarray}
among the four densities $n, N_1, N_2$ and $p$ only two are linearly independent.

The steady-state kinetics of paired defects is described by the $2I + 1$ equations
\begin{equation} \label{n2m}
\left( \dot{N}_{2,m} \right)_{\rm cb} + \left( \dot{N}_{2,m} \right)_{\rm vb} + \left( \dot{N}_{2,m} \right)_{\rm sr}= 0\:.
\end{equation}
The first term
\[
\left( \dot{N}_{2,m} \right)_{\rm cb} = 2 c_n \biggl( n_- \rho^{(m + \frac12)}_{\frac12, \frac12} +  n_+ \rho^{(m - \frac12)}_{-\frac12, -\frac12}\biggr)
\]
describes generation of the defect states with two electrons due to the capture of a conduction-band electron onto a single-electron defect. The second term
\[
\left( \dot{N}_{2,m} \right)_{\rm vb} = - c_p p N_{2,m}
\]
describes the recombination of a free unpolarized photohole with one of the singlet-state electrons, $c_p$ is the proportionality constant.
The final term describes the nuclear spin relaxation. For the nuclei with $I=1/2$ it has a simple unambiguous form
\begin{equation} \label{I12}
\left( \dot{N}_{2,m} \right)_{\rm sr} = - \frac{N_{2,m} - N_{2,-m}}{2 \tau^{(2)}_n} = - \frac{1}{\tau^{(2)}_n} \left( N_{2,m} - \frac{N_2}{2} \right)\:.
\end{equation}
In case of the nucleus $I=3/2$, the spin-relaxation term is ambiguous. However, if the perturbation leading to the inter-sublevel mixing is nonselective then, similarly to Eq.~(\ref{I12}), the relaxation for $I=3/2$ is characterized by one time parameter as follows \cite{DP1972}
\begin{equation} \label{rho12}
\left( \dot{N}_{2,m} \right)_{\rm sr} = - \frac{1}{\tau^{(2)}_n} \left( N_{2,m} - \frac{1}{2I+1} \sum\limits_{m'} N_{2,m'} \right) = - \frac{1}{\tau^{(2)}_n} \left( N_{2,m} - \frac{N_2}{2I+1} \right)\:.
\end{equation}

The kinetic equations for the spin-density matrices $\rho^{(M)}_{ss'} = \rho_{sm,s'm'}$ ($s + m = s' + m' = M$) can be written in the compact form as
\begin{equation} \label{rho1}
\dot{\hat{\rho}}^{(M)}_{\rm cb} +
\dot{\hat{\rho}}^{(M)}_{\rm vb}
+ \dot{\hat{\rho}}^{(M)}_{\rm esr} + \dot{\hat{\rho}}^{(M)}_{\rm nsr}= \frac{\rm i}{\hbar} \left[ {\cal H}^{(M)} \hat{\rho}^{(M)} \right]\:.
\end{equation}
In Eq.~(\ref{rho1}) the first and second terms
\begin{equation} \label{firsec}
\left(\dot{\rho}_{ss'}^{(M)}\right)_{\rm cb} = - c_n (n_{-s} + n_{-s'}) \rho_{ss'}^{(M)}\:,\:
\left(\dot{\rho}^{(M)}_{ss'}\right)_{\rm vb} = \frac{c_p}{2} N_{2,M-s} p \delta_{ss'}
\end{equation}
describe the capture and loss of the second electron by a defect. The term on the right-hand side represents the hyperfine and Zeeman interactions with a 2$\times$2 $M$-dependent spin Hamiltonian
\[
{\cal H}^{(M)} = \hbar (u_M s_z + v_M s_x)\:,\: u_M = \beta + M \Omega \:,\: v_M = \Omega \sqrt{\left( I + \frac12 \right)^2 - M^2}\:,
\]
where $\Omega = A/\hbar$, $\beta = g \mu_B B_z/\hbar$, $s_{\alpha} = \sigma_{\alpha}/2$ and $\sigma_{\alpha}$ are the spin Pauli matrices.
The bound-electron spin relaxation is phenomenologically described by the standard term
\[
\left( \dot{\rho}_{sm;s' m'} \right)_{\rm esr} = - \frac{1}{\tau_{sc}} \left(  \rho_{sm;s' m'} - \frac{\delta_{ss'}}{2} \sum\limits_{s''} \rho_{s''m;s'' m}\right)
\]
which is equivalent to
\begin{equation}
\left( \dot{\rho}^{(M)}_{ss'} \right)_{\rm esr} = - \frac{1}{\tau_{sc}} \left[  \rho^{(M)}_{ss'} - \frac{\delta_{ss'}}{2} \biggl( \rho^{(M)}_{ss} + \rho^{(M - 2s)}_{-s,-s} \biggr) \right]\:.
\end{equation}
Similarly to Eqs.~(\ref{I12}) and (\ref{rho12}), the nuclear spin relaxation can simply be described by
\begin{equation} \label{rho2a}
\left( \dot{\rho}_{sm,s' m'} \right)_{\rm nsr}  = - \frac{1}{\tau^{(1)}_{n}} \left(  \rho_{sm;s' m'} - \frac{\delta_{mm'}}{2I+ 1} \sum\limits_{m''} \rho_{sm'',sm''} \right)
\:,
\end{equation}
or, see Eq.~(\ref{rhoe}),
\begin{equation} \label{rho2}
\left( \dot{\rho}^{(M)}_{ss'} \right)_{\rm nsr}  = - \frac{1}{\tau^{(1)}_{n}} \left(  \rho^{(M)}_{ss'} - \delta_{ss'} \frac{\rho^e_{ss}}{2I+ 1} \right) \:.
\end{equation}
We remind that, for nonzero density-matrix components, the sum $s+m$ coincides with $s' + m'$ which means that $s=s'$  if $m=m'$. The set of equations (\ref{rho1}) represents $8I + 2$ scalar equations, particularly, 6 equations for $I=1/2$ and 14 equations for $I=3/2$.

The summation of the terms in Eq.~(\ref{rho1}) over $M$ yields the equations for the densities $N_{\pm}$ of single-electron defects, see Eqs.~(\ref{1a}) and (\ref{1b}),
\begin{eqnarray} \label{twosecond}
&&2c_n N_+ n_- + \frac{N_+ - N_-}{2 \tau_{sc}} + \frac{\rm i}{\hbar} \sum\limits_M \left[ {\cal H}^{(M)} \hat{\rho}^{(M)} \right]_{\frac12, \frac12}= 0\:, \label{2a}\\
&&2c_n N_- n_+ + \frac{N_- - N_+}{2 \tau_{sc}} + \frac{\rm i}{\hbar} \sum\limits_M \left[ {\cal H}^{(M)} \hat{\rho}^{(M)} \right]_{-\frac12, -\frac12} = 0 \label{2b}\:. \nonumber
\end{eqnarray}

The off-diagonal components of the spin-density matrix $\hat{\rho}^{(M)}$ can be expressed via the diagonal components
\begin{equation} \label{offdiag}
\rho_{\frac12, -\frac12}^{(M)}   = \frac{\rm i}{\hbar} \frac{{\cal H}^{(M)}_{\frac12,-\frac12} \left( \rho_{\frac12, \frac12}^{(M)} - \rho_{-\frac12, - \frac12}^{(M)} \right)}{c_n n + \frac{1}{\tau_{sc}} + \frac{1}{\tau^{(1)}_n} + \frac{\rm i}{\hbar} \left( {\cal H}^{(M)}_{\frac12,\frac12} - {\cal H}^{(M)}_{-\frac12,-\frac12} \right)}\:.
\end{equation}
Excluding the off-diagonal components we obtain for the diagonal components of the commutator in Eq.~(\ref{rho1})
\begin{equation} \label{commu2}
\frac{\rm i}{\hbar} \left[ {\cal H}^{(M)} \hat{\rho}^{(M)} \right]_{\frac12, \frac12} = \frac{\rm i}{\hbar} \left( {\cal H}^{(M)}_{\frac12, - \frac12} \rho_{-\frac12, \frac12}^{(M)}
- \rho_{\frac12, - \frac12}^{(M)}  {\cal H}^{(M)}_{- \frac12, \frac12}\right) = U_M \left( \rho_{\frac12, \frac12}^{(M)} - \rho_{-\frac12, - \frac12}^{(M)}\right)\:,
\end{equation}
where
\begin{equation} \label{commu+}
U_M = \frac{v_M^2 \tau_c}{2} \frac{1 + \tau_c/\tau_{sc} + \tau_c/\tau_n^{(1)} }{\left( 1 + \tau_c/\tau_{sc} + \tau_c/\tau_n^{(1)} \right)^2 + u_M^2 \tau_c^2}
\end{equation}
and
\begin{equation} \label{tauc}
\tau_c = \frac{1}{c_n n} \:.
\end{equation}
The factor $U_0$ is an even function of the longitudinal magnetic
field whereas the factors $U_M$ with $M \neq 0$ are asymmetric
functions of $B_z$ because
\begin{equation} \label{betao}
u^2_M = (\beta + M \Omega)^2 = \left( g \mu_B B_z + M A \right)^2/\hbar^2\:.
\end{equation}
Under circularly-polarized photoexcitation the electron-nuclear states with $M$ and $-M$ ($M \neq 0$) can be differently involved in the kinetics which is the main reason for the asymmetry of dependence
$P_e(B_z)$ observed experimentally.

The expression (\ref{commu2}) describing the effect of hyperfine interaction can be rewritten in the form
\begin{equation} \label{golden}
\frac{2 \pi}{\hbar} \left( \frac{v_{M}}{2} \right)^2 \left( \rho^{(M)}_{\frac12,\frac12} - \rho^{(M)}_{-\frac12,-\frac12}\right)  \frac{1}{\pi} \frac{\hbar/ \tilde{\tau}_c}{\left( \hbar u_{M} \right)^2 + \left( \hbar / \tilde{\tau}_c \right)^{2}} \:,
\end{equation}
allowing the interpretation in the spirit of Fermi's golden rule for the probability rate
\[
W_{21} = \frac{2 \pi}{\hbar} |V_{21}|^2 (f_1 - f_2) \delta (E_2 - E_1)
\]
of the transition from the quantum state $|1\rangle$ to the state $|2\rangle$, where $E_2$ and $ E_1$ are the energies of these states, $f_2$ and $f_1$ are their average occupations, $V_{21}$ is the matrix element of the perturbation operator. In Eq.~(\ref{golden}), the role of ideal $\delta$-function is played by the smoothed $\delta$-function with the damping
\begin{equation} \label{tildetau}
\frac{1}{\tilde{\tau}_c} = \frac{1}{\tau_c} + \frac{1}{\tau_{sc}} + \frac{1}{\tau_n^{(1)} }\:.
\end{equation}

\subsection{The model neglecting nuclear spin relaxation}
If the nuclear spin relaxation is neglected then the set of kinetic equations reads
\begin{eqnarray} \label{n2m32}
&&c_p N_{2,m} p = 2 c_n \left( n_- \rho_{\frac12,\frac12}^{(m + \frac12)} + n_+ \rho_{-\frac12,-\frac12}^{(m - \frac12)}\right)\:, \\
&&2 c_n n_{-s} \rho_{ss}^{(M)} + U'_M \left( \rho_{ss}^{(M)} - \rho_{-s,-s}^{(M)} \right) + \frac{1}{2 \tau_{sc}} \left( \rho_{ss}^{(M)}  - \rho^{(M-2s)}_{-s,-s} \right)= \frac{c_p}{2} N_{2,M - s} p
\:, \nonumber
\end{eqnarray}
where
\[
U'_M  = \frac{1}{2 \tau_c} \frac{v_M^2 \tau_c^2[1 + (\tau_c/\tau_{sc})]}{[1 + (\tau_c/\tau_{sc})]^2 + u_M^2 \tau_c^2} \:.
\]
We remind that, for $M = \pm( I + 1/2)$,  the value of $u_M$ vanish and, thus, $U'_M$ is nonzero only for $|M| < I + 1/2$.

Surprisingly, the set (\ref{n2m32}) has a simple magnetic-field-independent solution
\begin{eqnarray} \label{solut}
&&P_e = \frac{n_+ - n_-}{n_+ + n_-} = \frac{P_i G T}{1 - \eta}\:,\: P_c = \frac{N_+ - N_-}{N_+ + N_-} = \frac{T_c}{\tau_c}P_e\:,\\
&& \rho^{(M)}_{ss} = C_I N_1 (1 + P_c)^{J+M}(1 - P_c)^{J-M}\:, \nonumber\\
&& N_{2,m} = 2 C_I N_2 (1 + P_c)^{I+m}(1 - P_c)^{I-m}\:, \nonumber
\end{eqnarray}
where $G = G_+ + G_-$ is the total optical generation rate of photoelectrons into the conduction band (or, equivalently, photoholes into the valence band),
$P_i = (G_+ - G_-)/(G_+ + G_-)$ is the initial degree of photoelectron spin polarization,
\[
\eta = \frac{TT_c}{\tau \tau_c}\:,\:\tau = \frac{1}{c_nN_1}\:,\: \frac{1}{T} = \frac{1}{\tau_s} + \frac{1}{\tau} \:,\: \frac{1}{T_c} = \frac{1}{\tau_{sc}} + \frac{1}{\tau_c}\:,
\]
and the time $\tau_c$ is defined by Eq.~(\ref{tauc}). The factor $C_I$ is given by
\[
C_I = \frac{P_c}{(1+P_c)^{2I + 1} - (1-P_c)^{2I + 1}}
\]
and equals to 1/4 for $I=1/2$ and to $[8(1 + P_c^2)]^{-1}$ for $I=3/2$.

One can see that, for the steady-state solution (\ref{solut}), the values $\rho_{ss}^{(M)}$ and $\rho_{-s, -s}^{(M)}$ coincide. This means that, on the first hand, the diagonal components of the commutator in Eqs.~(\ref{2a}), (\ref{2b}) and (\ref{commu2}) are switched off as if the hyperfine interaction were absent and, on the other hand, the nuclei are spin-polarized and their spin polarizations in the single- and double-electron defect states coincide
\[
\frac{\sum\limits_s \rho_{sm,sm}}{N_1} = \frac{N_{2,m}}{N_2} \:.
\]

Since in the steady state the term $U'_M \left( \rho_{ss}^{(M)} - \rho_{-s,-s}^{(M)} \right)$ in Eq.~(\ref{n2m32}) vanishes the densities of conduction-band electrons, $n$, and double-electron defects, $N_2$, satisfy equations independent of the hyperfine constant
$A$ and the magnetic field:
\begin{eqnarray} \label{XYZequation}
&&\mbox{}\hspace{3 cm}Y(Y+Z) = X \:,\\ &&\frac{1-Y}{a} \left\{ Z - P_i^2 \left(
\frac{\tau_s}{\tau_h^*} \right)^2 \frac{X^2 \left( Z +
\tau^*/\tau_{sc} \right)}{[Z + \tau^*/\tau_{sc} + (1-Y)
\tau_s/\tau_{sc}]^2} \right\} = X\:,  \nonumber
\end{eqnarray}
where $\tau^* = (c_n N_c)^{-1}$, $\tau^*_h = (c_p N_c)^{-1}$, $a = c_p / c_n$ and we use the dimensionless variables
\begin{equation} \label{XYZa}
X = \frac{G}{c_p N_c^2}\:,\: ~~~~Y = \frac{N_2}{N_c} = \frac{N_c -
N_1}{N_c} \:,\: ~~~~Z = \frac{n}{N_c}\:.
\end{equation}
In these notations the hole density $p$ is given by $(Y + Z) N_c$. Equations (\ref{XYZequation}) are identical to Eqs.~(20) in Ref.~\cite{Condensed} derived neglecting electron-nuclear hyperfine interaction.
\subsection{The model assuming fast nuclear spin relaxation in the paired singlet}
As an alternative limiting case, we assume the nuclear spin relaxation in the defect state with two electrons to be quite short and set
\[
N_{2,m} = \frac{N_2}{2I + 1}
\]
in Eq.~(\ref{n2m32}) and, similarly to \cite{DP1972}, ignore the nuclear spin relaxation in the single-electron defects. For convenience, we will first ignore the spin relaxation of bound electrons and then will extend the obtained result to allow for this relaxation. The solution for the spin-density matrix can be presented in the form
 \begin{equation} \label{taun0}
\rho^{(M)}_{\pm \frac12, \pm \frac12} = \frac{c_p p N_2}{2(2I + 1) c_n n} \frac{ ( 1 \pm P_e ) (1 + u_M^2 \tau_c^2) + v_M^2 \tau_c^2}{( 1 - P^2_e ) (1 + u_M^2 \tau_c^2) +  v_M^2 \tau_c^2 }   \:.
\end{equation}
Note that since $v_{\pm \left( I + 1/2\right)} = 0$ the components $\rho^{(I + 1/2)}_{ss}$ and $\rho^{(-I - 1/2)}_{ss}$ reduce to a much simpler form
\[
\rho^{(I+1/2)}_{\frac12, \frac12} = \frac{c_p p N_2}{2(2I + 1) c_n n (1 - P_e)}\:,\:\rho^{(-I-1/2)}_{- \frac12, - \frac12} = \frac{c_p p N_2}{2(2I + 1) c_n n (1 + P_e)}\:.
\]
Moreover, the unphysical states with $s=1/2, M = -(I+1/2)$ and $s=-1/2, M = I+1/2$ should be excluded from Eq.~(\ref{taun0}).

Using the identity
\[
\frac{ ( 1 \pm P_e ) (1 + u_M^2 \tau_c^2) + v_M^2 \tau_c^2}{( 1 - P^2_e ) (1 + u_M^2 \tau_c^2) +  v_M^2 \tau_c^2 }
= \frac{1}{1 \mp P_e} \left( 1 \mp  \rho_e \frac{v_M^2 \tau_c^2}{( 1 - P^2_e ) (1 + u_M^2 \tau_c^2) +  v_M^2 \tau_c^2 }\right)\:,
\]
we derive for the densities $N_+$, $N_-$ and the polarization degree $P_c$ the following expressions
\begin{equation} \label{npm}
N_+ = \frac{1 - \zeta P_e}{2(1 - P_e)} G \tau_c\:, \:
N_- = \frac{1 + \zeta P_e}{2(1 + P_e)} G \tau_c\:,\:P_c = \frac{1 - P_e}{1 - \zeta P_e} P_e\:,
\end{equation}
where
\begin{equation} \label{zeta}
\zeta = \frac{1}{2I + 1} \sum\limits_M  \frac{ v_M^2 \tau_c^2}{( 1 - P^2_e ) (1 + u_M^2 \tau_c^2) +  v_M^2 \tau_c^2 }\:.
\end{equation}
Replacing $N_{\pm}$ by their expressions (\ref{npm}) we find
\begin{equation} \label{petilde}
P_e = \frac{P_i G \tilde{\tau}_s}{n} = \frac{P_i X \tilde{\tau}_s}{Z \tau^*}\:,
\end{equation}
where
\[
 \frac{1}{\tilde{\tau}_s} = \frac{1}{\tau_s} + \zeta \frac{G}{n}\:.
\]
The densities $n$ and $N_2$ satisfy Eqs.~(\ref{XYZequation}) where
\[
\tau_s^2 \frac{Z + \tau^*/\tau_{sc}}{[Z + \tau^*/\tau_{sc} + (1-Y) \tau_s/\tau_{sc}]^2}
\]
should be replaced by
\begin{equation} \label{replace}
{\tau_s^*}^2 = \frac{1 - \zeta}{1 - \zeta P_e^2} \frac{\tilde{\tau}_s^2}{Z} \:.
\end{equation}

The spin relaxation of bound electrons can easily be incorporated into the balance equations if $\tau_{sc} \gg \tau_c$. For this purpose the sum in Eqs.~(\ref{2a}) and (\ref{2b})
can be approximated by the sum calculated in the limit $\tau_{sc} \to \infty$ and given by $\zeta G P_e/2$.
As a result the problem is reduced to solving a set of four equations, namely, the two equations (\ref{1a}), (\ref{1b}) and two additional equations
\begin{eqnarray} \label{balancen2}
&&2c_n N_+ n_- + \frac{\zeta}{2} G P_e + \frac{N_+ - N_-}{2 \tau_{sc}}= \frac{c_p}{2}N_2 p\:, \\
&&2c_n N_- n_+ - \frac{\zeta}{2} G P_e + \frac{N_- - N_+}{2 \tau_{sc}}= \frac{c_p}{2}N_2 p\: \nonumber\:.
\end{eqnarray}
It follows then that Eqs.~(\ref{petilde}) and (\ref{replace}) are valid as well if $\tilde{\tau}^{-1}_s$ is replaced by
\begin{equation} \label{tau*2}
\frac{1}{\tilde{\tau}_s} = \frac{1}{\tau_s} + \zeta \frac{G}{n} + \frac{N_1}{n} \frac{1}{\tau_{sc}} = \frac{1}{\tau_s} + \frac{\zeta X}{\tau^*_h Z} + \frac{1-Y}{\tau_{sc} Z} \:.
\end{equation}

One can see from Eqs.~(\ref{replace}) and (\ref{tau*2}) that, for the fast spin relaxation of double-electron defect states, the hyperfine interaction effectively leads to a decrease of the electron spin relaxation time governed by the parameter $\zeta$. Since $\zeta$ is an even function of $B_z$, see Eq.~(\ref{zeta}), in the approximation under consideration the point of minimum in the dependence $P_e(B_z)$ lies at $B_z = 0$.
\subsection{Approximation of unpolarized nuclei}
At low excitation powers when the lifetime of single-electron defect state $\tau_c = (c_n n)^{-1}$ is long compared with $\tau^{(1)}_n$ and that of two-electron states $\tau_{c2} = (c_p p)^{-1}$ is longer than $\tau^{(2)}_n$ one can take the nuclei to be unpolarized and set
\[
\rho^{(M)}_{\pm \frac12, \pm \frac12 } = \frac{N_{\pm}}{2I + 1}\:.
\]
It follows then that the third terms describing in Eqs.~(\ref{twosecond}) the hyperfine interaction can be replaced by
\[
\frac{\rm i}{\hbar} \sum\limits_M \left[ {\cal H}^{(M)} \hat{\rho}^{(M)} \right]_{\frac12, \frac12} = - \frac{\rm i}{\hbar} \sum\limits_M \left[ {\cal H}^{(M)} \hat{\rho}^{(M)} \right]_{-\frac12, -\frac12} = \frac{N_+ - N_-}{2 \tau_{scn}}\:,
\]
where $\tau^{-1}_{scn}$ is the bound-electron spin relaxation
rate induced by the nucleus and defined by
\[
\frac{1}{\tau_{scn}} = \frac{2}{2 I + 1} \sum\limits_M U_{M}\:.
\]
Therefore, in this approximation the influence of nuclei is accounted for by replacing $\tau_{sc}^{-1}$ by the sum $\tau_{sc}^{-1} + \tau_{scn}^{-1}$.

\section{Hyperfine interaction for a nucleus with $I = 1/2$}
In this case Eq.~(\ref{n2m}) reduces to two scalar equations for $N_{2,1/2}$ and $N_{2,-1/2}$ which can be transformed to the equations for $N_{2,1/2}$ and $N_2 = N_{2,1/2} + N_{2,-1/2}$:
\begin{eqnarray} \label{n2ma}
&& c_p p N_{2,\frac12} + \frac{1}{\tau^{(2)}_n} \left( N_{2,\frac12} - \frac{N_2}{2} \right) = 2 c_n \left( n_- \rho_{\frac12, \frac12}^{(1)} + n_+ \rho_{-\frac12, - \frac12}^{(0)} \right) \:, \\
&& c_p p N_2 = G\:. \nonumber
\end{eqnarray}
Equations~(\ref{rho1}) for $M=0$ read
\begin{eqnarray}
&&2 c_n n_- \rho_{\frac12,\frac12}^{(0)} + \frac{\rm i}{\hbar} \left[ {\cal H}^{(0)} \hat{\rho}^{(0)} \right]_{\frac12, \frac12} + \frac{\rho_{\frac12, \frac12}^{(0)} - \rho_{-\frac12, - \frac12}^{(-1)}}{2 \tau_{sc}} + \frac{\rho_{\frac12, \frac12}^{(0)} - \rho_{\frac12, \frac12}^{(1)}}{2 \tau_n^{(1)}} = \frac{c_p}{2} N_{2, - \frac12} p \:, \label{12120}\\
&&2 c_n n_+ \rho_{-\frac12,-\frac12}^{(0)} + \frac{\rm i}{\hbar} \left[ {\cal H}^{(0)} \hat{\rho}^{(0)} \right]_{-\frac12, -\frac12} + \frac{\rho_{-\frac12, -\frac12}^{(0)} - \rho_{\frac12, \frac12}^{(1)}}{2 \tau_{sc}} + \frac{\rho_{-\frac12, -\frac12}^{(0)} - \rho_{-\frac12, -\frac12}^{(1)}}{2 \tau_n^{(1)}} = \frac{c_p}{2} N_{2, \frac12} p \:, \nonumber
\end{eqnarray}
where, see Eq.~(\ref{commu2}),
\begin{equation} \label{commu}
\frac{\rm i}{\hbar} \left[ {\cal H}^{(0)} \hat{\rho}^{(0)} \right]_{\frac12, \frac12} = U_0 \left( \rho_{\frac12, \frac12}^{(0)} - \rho_{-\frac12, - \frac12}^{(0)}\right)\:.
\end{equation}
Two additional equations for $\rho_{\frac12,\frac12}^{(1)}$ and $\rho_{-\frac12,-\frac12}^{(-1)}$ have the form
\begin{eqnarray}
&&2 c_n n_- \rho_{\frac12,\frac12}^{(1)} + \frac{\rho_{\frac12, \frac12}^{(1)} - \rho_{-\frac12, -\frac12}^{(0)}}{2 \tau_{sc}} + \frac{\rho_{\frac12, \frac12}^{(1)} - \rho_{\frac12, \frac12}^{(0)}}{2 \tau_n^{(1)}} = \frac{c_p}{2} N_{2, \frac12} p \:, \label{ss1} \\
&&2 c_n n_+ \rho_{-\frac12,-\frac12}^{(-1)} + \frac{\rho_{-\frac12, -\frac12}^{(-1)} - \rho_{\frac12, \frac12}^{(0)}}{2 \tau_{sc}} + \frac{\rho_{-\frac12, -\frac12}^{(-1)} - \rho_{-\frac12, -\frac12}^{(0)}}{2 \tau_n^{(1)}} = \frac{c_p}{2} N_{2, -\frac12} p \:.\nonumber
\end{eqnarray}

From Eqs.~(\ref{12120}) and (\ref{commu}) we conclude that the set of equations for the diagonal components of the spin-density matrix and occupations $N_{2,m}$ are dependent on the magnetic field through the square $u_0^2 = (g \mu_B B_z \tau_c/ \hbar)^2$. This clearly demonstrates that, for $I = 1/2$, the electron spin polarization $P_e = (n_+ - n_-)/(n_+ + n_-)$ is a symmetric function of $B_z$ and has a minimum at the point $B_z=0$.

Equations (\ref{1a})$-$(\ref{densities2}), (\ref{n2ma})$-$(\ref{ss1}) form a complete set to be solved. It may be further converted for a more convenient numerical calculation. By using Eqs.~(\ref{12120}) and (\ref{ss1}) we can establish a linear relation
\begin{equation} \label{rcl5}
 {\rho}_{s,s}^{(M)}  = \frac{c_p p}{2 c_n n} \sum\limits_{m = \pm 1/2} Q^{(M)}_{s,m}N_{2,m}
 \end{equation}
between the components of spin-density matrix ${\rho}_{ss}^{(M)}$ and the densities $N_{2,\pm 1/2}$. The expressions for dimensionless coefficients $Q^{(M)}_{s,m}$ have a simple but cumbersome form and are not presented here. Substituting (\ref{rcl5}) into Eq.~(\ref{n2ma}) we find $N_{2,\pm 1/2}$ and $f_{m}=N_{2,m}/N_{2}$.
The next step is to write down expressions for the densities of single-electron defects with the electron spin $\pm 1/2$, namely,
\begin{equation}\label{rc17}
N_{+}=\sum\limits_{M=0,1} \rho_{1/2,1/2}^{(M)}=N_{2}\frac{c_p p}{2c_n n} D_{1/2}\:,\: N_{-}=\sum\limits_{M=-1,0} \rho_{-1/2,-1/2}^{(M)}= N_{2}\frac{c_p p}{2c_n n} D_{-1/2}\:,
\end{equation}
where the coefficients
\begin{equation}\label{rc18}
D_{1/2}=\sum\limits_{m} \sum\limits_{M=0,1} Q^{(M)}_{1/2,m} f_{m}\:,\: D_{-1/2}=\sum\limits_{m} \sum\limits_{M=-1,0} Q^{(M)}_{-1/2,m} f_{m}
\end{equation}
are functions of $n_{+}$, $n_{-}$ and parameters of the model.

The substitution of (\ref{densities2}), (\ref{rc17}) into Eqs.~(\ref{1a}), (\ref{1b}) gives us two equations connecting three unknown quantities $n_{+}$, $n_{-}$ and $N_2$. We replace them by their sum and difference and obtain
\begin{eqnarray}
&&Y(Y+Z) = X \:, \label{rcl24} \\
&&XP_{i}-\frac{\tau^*_{h}}{\tau_{s}}P_{e}Z=X F \left(P_{e},Z \right)\:, \label{rcl25}
\end{eqnarray}
where
\[
F\left(P_{e},Z \right) =\frac{1+P_{e}}{2} D_{-1/2}\left(P_{e},Z \right)-\frac{1-P_{e}}{2} D_{+1/2}\left(P_{e},Z \right)\:,
\]
and the dimensionless variables (\ref{XYZa}) are used. Note that Eq.~(\ref{rcl24}) follows immediately from the second equation (\ref{n2ma}) and coincides with the first equation (\ref{XYZequation}). To find the third equation we express $N_{+}$, $N_{-}$ via $n_{\pm}$ by using Eqs.~(\ref{1a}), (\ref{1b}) and insert the expressions into Eq.~(\ref{densities2}) arriving at
\begin{equation} \label{rcl26}
X\left(1-P_{i}P_{e}\right)+
\frac{\tau_{h}^{*}}{\tau_{s}}P_{e}^2 Z=\frac{1-Y}{a}\left(1-P_{e}^2 \right) Z \:,
\end{equation}
where $a = c_p/c_n$ and hereafter, instead of $n_{\pm}$, we use the variables $n = n_+ + n_-$ and $P_{e} = (n_+ - n_-)/(n_+ + n_-)$.

From Eq.~(\ref{rcl26}) one has
\begin{equation} \label{rcl27}
Y=1-a\frac{ X \left( 1 - P_{i}P_{e} \right) + \left(\tau_{h}^{*}/\tau_{s}\right)Z P_{e}^2 }{Z \left( 1-P_{e}^2 \right)}=\frac{L + M Z}{Z}\:,
\end{equation}
where $L,M$ are independent of $Z$. Therefore the substitution of (\ref{rcl27}) into Eq.~(\ref{rcl24}) gives a third-order equation for $Z$. Two of the three solutions of this equation are positive and define the dependencies of electron concentration on $P_e$ and $X$. These dependencies together with (\ref{rcl25}) allow to find two values of $P_e$ at fixed photogeneration rate $X$, but one of the corresponding values of the density of double-electron defects $Y$ turns out to be negative, and the solution of Eqs.~(\ref{rcl24})$-$(\ref{rcl26}) is unique.

If the nuclear spin relaxation is neglected or the hyperfine coupling is completely absent then Eqs.~(\ref{rcl25}) and (\ref{rcl26}) reduce, respectively, to the first equation (\ref{solut}) and the second equation (\ref{XYZequation}).
\section{Results of computer calculation and discussion}

\begin{figure}
\includegraphics[width=14.0cm,angle=0]{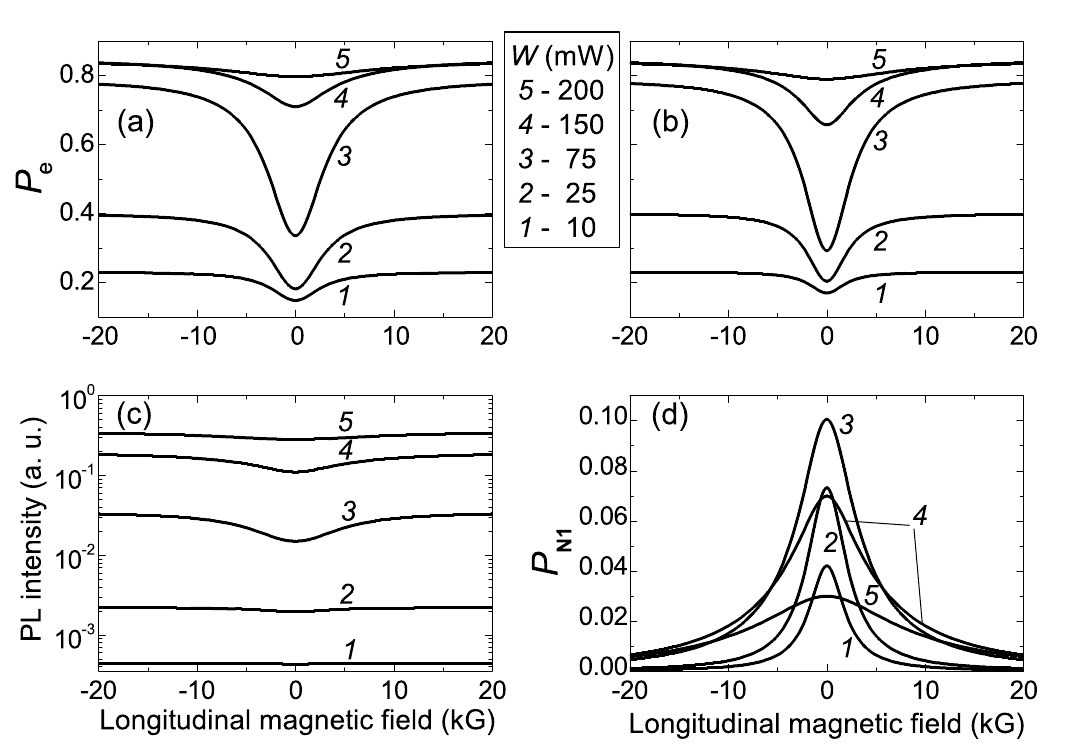}
\caption{\label{fig1} Spin polarization degree of the
conduction-band electrons (a,b), the intensity of interband
PL (c) and nuclear spin polarization (d) as a
function of the longitudinal magnetic field calculated for
different excitation powers $W$ of the circularly polarized light. The
curves 1 -- 5 correspond respectively to the following values of
$W$: 10, 25, 75, 150, and 200\,mW. The set of model
parameters $\tau^* = 2$\,ps, $\tau^*_h = 30$\,ps, $\tau_s =
140$\,ps, $\tau_{sc} = 700$\,ps, $P_i = 0.13$, $N_c = 3 \times
10^{15}$\,cm$^{-3}$, $A = 17$ $\mu$eV is used in the calculation.
The relation $G = 2 \times 10^{23} W$ between the generation rate and excitation
power is derived from the experiment, the units for $G$ and $W$ are cm$^{-3}$s$^{-1}$ and mW, respectively.
Panel (a): Both the nuclear spin relaxation time of the
one-electron defect state ($\tau_n^{(1)}$) and that of the
two-electron state ($\tau_n^{(2)}$) are taken to equal 150\,ps.
Panels (b), (c) and (d): $\tau_n^{(1)}$ = 1000\,ps, $\tau_n^{(2)}$
= 1\,ps. Surprisingly, the sets (a) and (b) of
the calculated curves are qualitatively similar. }
\end{figure}

The signature of the electron-nuclear hyperfine coupling in
single-electron defect states is a growth of the spin
polarization $P_e$ of conduction-band electrons and the interband
PL intensity $J$ with increasing the longitudinal
magnetic field as shown in Figs.~1(a), 1(b) and 1(c). In experiment this is
observed, under circularly polarized interband optical excitation,
via the magnetic-field induced increase in the PL circular
polarization and intensity \cite{KalevichPRB,JetpLett2012,Buyanova2013,NatureComm,Toulouse2014}.
The set of parameters unrelated to the nuclei and hyperfine
coupling is the same as used in the previous
analysis~\cite{KalevichPRB}:  $\tau^*$= 2 ps, $\tau^*_h$ = 30 ps, $\tau_s$ = 140 ps, $\tau_{sc}$ = 700 ps, $P_i$ = 0.13, $N_c$ = 3$\times$10$^{15}$ cm$^{-3}$. For the deep centers in
GaAsN the average hyperfine constant $A$ was estimated
as 6.9$\times$10$^{-2}$\;cm$^{-1}$ = 8.5\;$\mu$eV~\cite{naturemat,Chen2014}. For the nucleus $I=3/2$, the hyperfine splitting of
the states with the total angular momenta 2 and 1 equals $2A$. To
have a comparable strength of the hyperfine interaction with the
nucleus $I=1/2$ we take $A$ = 17\;$\mu$eV. The choice of the
nuclear spin relaxation time in the single-electron defect state,
$\tau_{n}^{(1)}$, causes the greatest difficulties. Apparently,
this phenomenological time parameter cannot be shorter than the
time $\tau_{n}^{(2)}$ describing spin relaxation of the nuclei in
the two-electron defect states. The growth of the polarization
$P_e$ illustrated in Fig.~1 is calculated for (a) the coinciding
times $\tau_{n}^{(1)}$ and $\tau_{n}^{(2)}$ and (b) for the short
time $\tau_{n}^{(2)}$ and long time $\tau_{n}^{(1)}$. The nuclear
spin polarization is characterized by the two polarization degrees
\[
P_{N1} = \frac{1}{N_1} \sum\limits_s \left( \rho_{s, \frac12; s,
\frac12} - \rho_{s, -\frac12; s, -\frac12}\right)\:,\:P_{N2} =
\frac{N_{2,\frac12} - N_{2,-\frac12}}{N_2} \:.
\]

\begin{figure}
\includegraphics[width=7.0cm,angle=0]{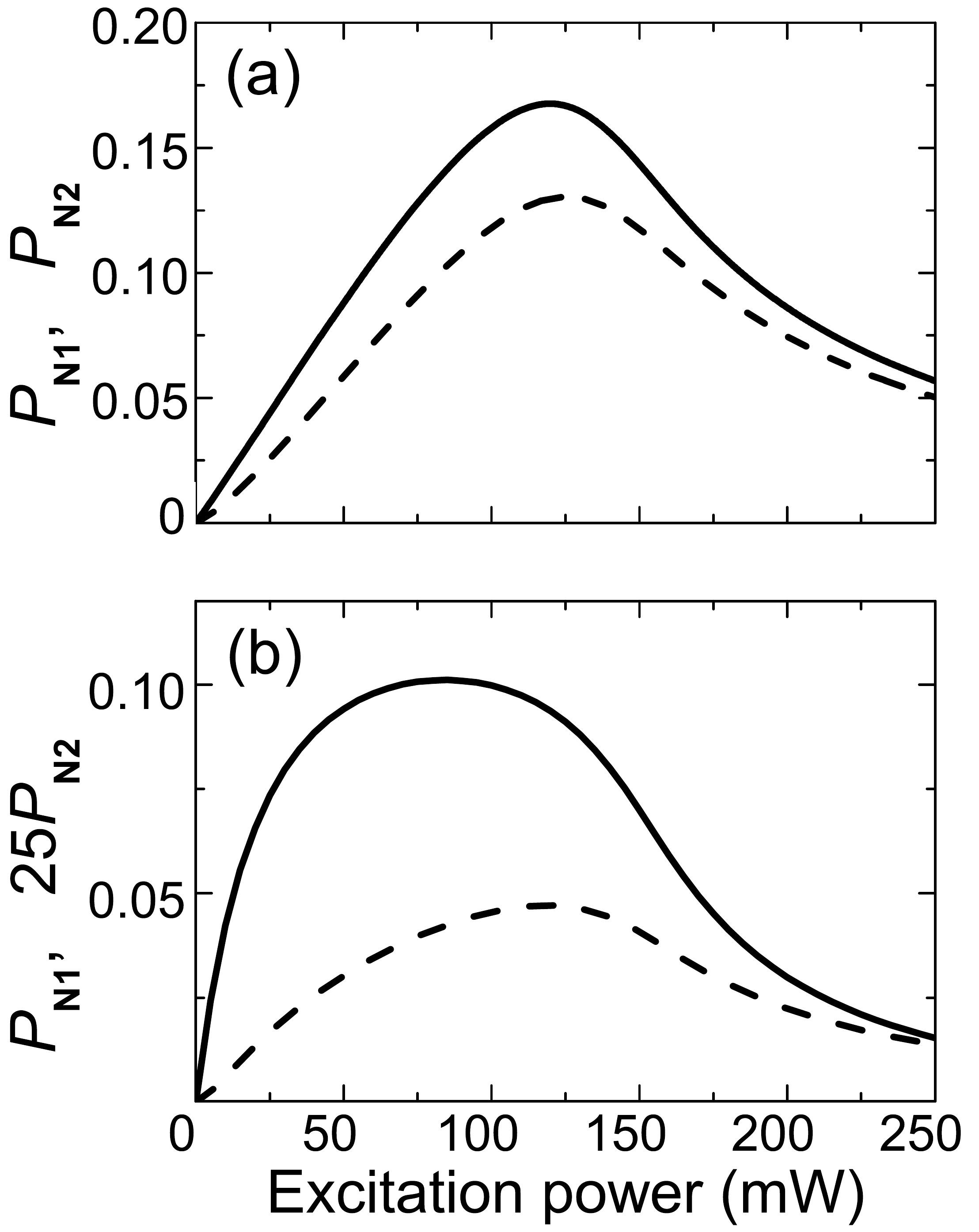}
\caption{\label{fig2} Nuclear spin polarizations $P_{N1}$ (solid
curves) and $P_{N2}$ (dashed) as a function of the excitation
power calculated at zero magnetic field for (a) $\tau_n^{(1)} =
\tau_n^{(2)} = 150$\,ps and (b) $\tau_n^{(1)}$ = 1000\,ps,
$\tau_n^{(2)} = 1$\,ps. In panel~(b) the values of $P_{N2}$ are
multiplied by the factor of 25.}
\end{figure}

Their dependence on the excitation power $W$ calculated in the absence
of magnetic field is depicted in Fig.~2. In the case (a) the
values $P_{N1}$ and $P_{N2}$ are different but comparable in
magnitude whereas in the case (b), as expected, the polarization
$P_{N2}$ is small and the polarization $P_{N1}$ is of the same
order as the polarizations in Fig.~2(a). As seen in Fig.~1(d), the
average nuclear spin is the highest at $B=0$ and exhibits
depolarization with the increasing magnetic field since the field
decouples the electron and nuclear spins. It is also worth to note
that the zero-field nuclear spin polarization is a nonmonotonic
function of the excitation power and reaches a maximum for the intermediate
power $\sim$75 mW. This can be understood as follows: in the low-power limit, the system is slightly driven out of the equilibrium
and the nuclear polarization is still weak; in the high-power limit, the lifetime of bound electrons $\tau_c$ is very short, the
hyperfine-coupling factor $U_M$ in Eqs.~(\ref{commu2}), (\ref{commu+}) decreases and the dynamic nuclear polarization by the polarized
electrons is strongly weakened.

\begin{figure}
\includegraphics[width=7.0cm,angle=0]{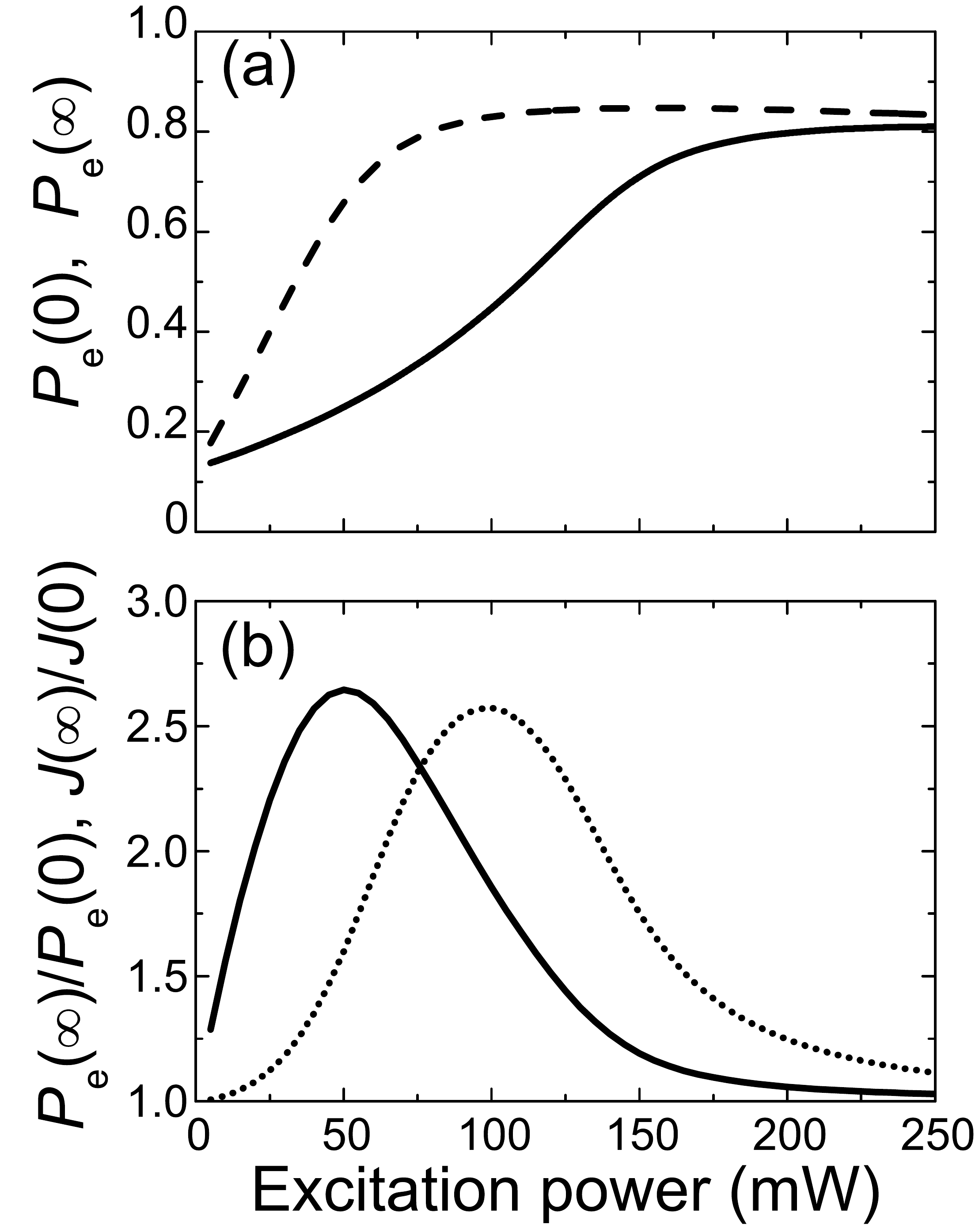}
\caption{\label{fig3} (a) Spin polarization of the conduction-band
electrons as a function of the excitation power calculated at zero
magnetic field (solid curve) and infinitely high magnetic field
(dashed curve). The parameters are the same as in Fig. 1(b). (b)
The dependence of ratios $P_e(\infty)/P_e(0)$ (solid) and
$J(\infty)/J(0)$ (dotted) on the excitation power.}
\end{figure}

The spin-filtering effect is demonstrated in Fig.~3(a). At very
low excitation intensity this effect is not switched on, the
degree $P_e$ is independent of the magnetic field and the two
curves in the figure calculated at zero (solid) and infinitely
high (dashed) magnetic field merge as $W \to 0$. At very high
pumping the curves again merge since the lifetime
$\tilde{\tau}_c$, see Eq.~(\ref{tildetau}), becomes very short and
the uncertainty caused by this reduction decouples the hyperfine
interaction. Figure 3(b) shows the power dependence of the ratio
of electron polarizations at the strong and zero magnetic fields
(solid) and similar ratio of the PL intensities (dotted). The
peaks of the two curves are shifted with respect to each other in
agreement with the experiment, Fig.~4(b) in
Ref.~\cite{KalevichPRB}.

Figure 4 illustrates the sensitivity of the polarization $P_e$ to
variation of the nuclear spin relaxation time in the models with
$\tau_n^{(1)} = \tau_n^{(2)}$ and $\tau_n^{(1)} \neq
\tau_n^{(2)}$. It is clear from the figure that there exists a
critical interval of the time values above which the electron
polarization ceases to depend on the magnetic field confirming the
conclusion of Sec. III A. The detailed calculation shows that
this interval lies around 1000 ps. On the other hand, at extremely
short nuclear spin relaxation times the magnetic field dependence
of $P_e$ disappears as well, due to the increasing uncertainty of
$\tilde{\tau}_c^{-1}$ in Eqs.~(\ref{golden}), (\ref{tildetau}).

\begin{figure}[t]
\includegraphics[width=7.0cm,angle=0]{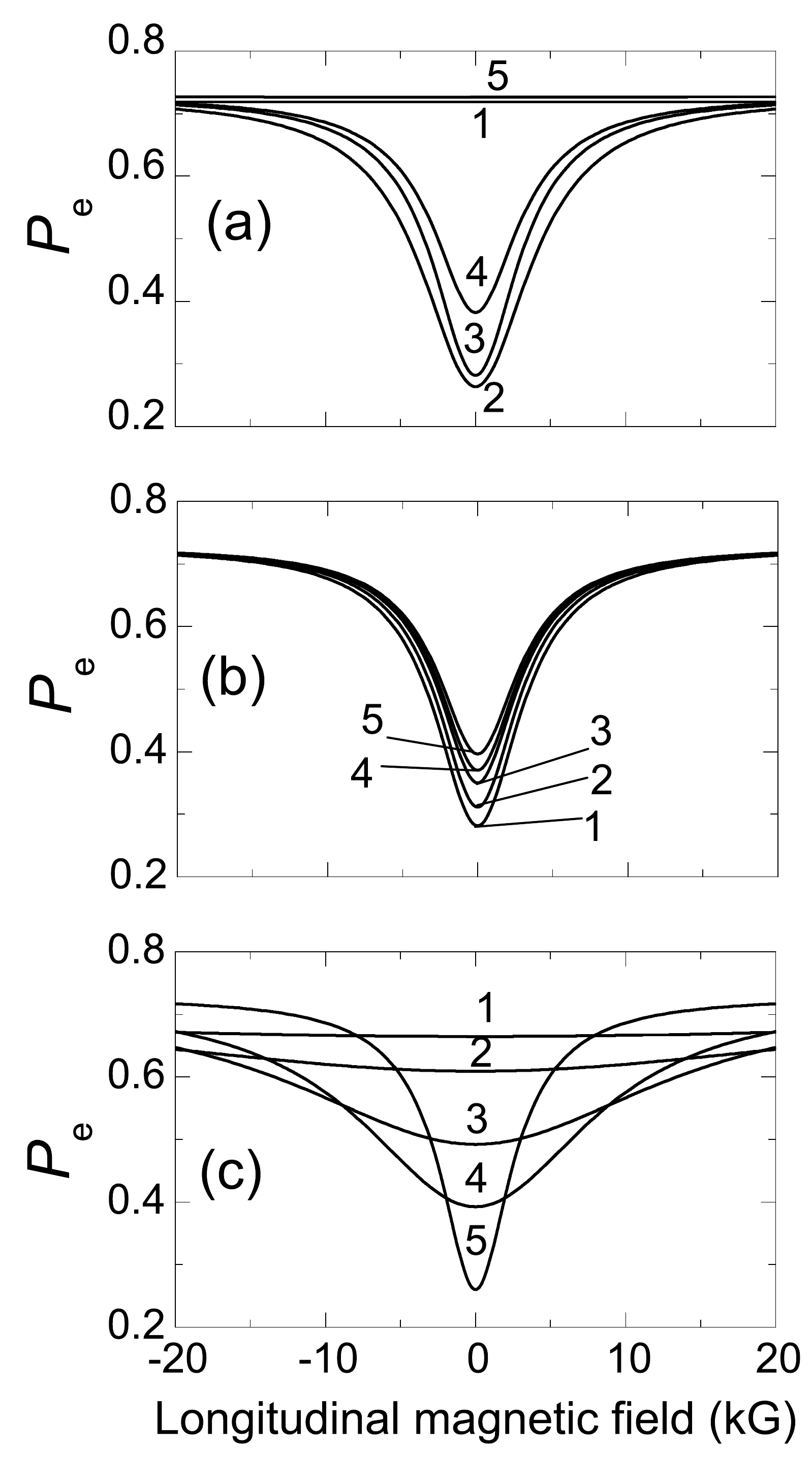}
\caption{\label{fig4} Dependence of the spin polarization of the
conduction-band electrons on the magnetic field calculated for the
excitation power 60\,mW and different values of the nuclear spin
relaxation time $\tau_n^{(1)}$ (a) assuming $\tau_n^{(1)}$ and
$\tau_n^{(2)}$ to coincide and being equal to $0.15$\,ps (curve
1), $50$\,ps (2), $150$\,ps (3),  $450$\,ps (4), and $150000$\,ps
(5); (b) keeping $\tau_n^{(2)}$ = 150\,ps fixed and $\tau_n^{(1)}$
taking the values 150\,ps (curve 1), 300\,ps (2), 750\,ps (3),
1500\,ps (4) and $150000$\,ps (5); and (c) $\tau_n^{(2)}$ = 1\,ps
and $\tau_n^{(1)}$ = 1\,ps (curve 1), 2\,ps (2), 5\,ps (3), 10\,ps
(4), and 1000\,ps (5). Other parameters are the same as in
Fig.~1(b). }
\end{figure}

Finally, Fig.~5 depicts the magnetic field dependence of $P_e$ for
different values of the hyperfine constant $A$. At zero $A$ the
electron polarization is insensitive to the longitudinal magnetic
field. With $A$ increasing up to 68\;$\mu$eV the zero-field value
of $P_e$ decreases by a factor of $\sim$3. The halfwidth $B_{\rm
hw}$ of the recovery curve
$$
\frac{P_e(B) - P_e(0)}{P_e(\infty) - P_e(0)}
$$
increases sublinearly and is more sensitive to the variation of $A$ as compared to $P_e(0)$.

\begin{figure}
\includegraphics[width=8.0cm,angle=0]{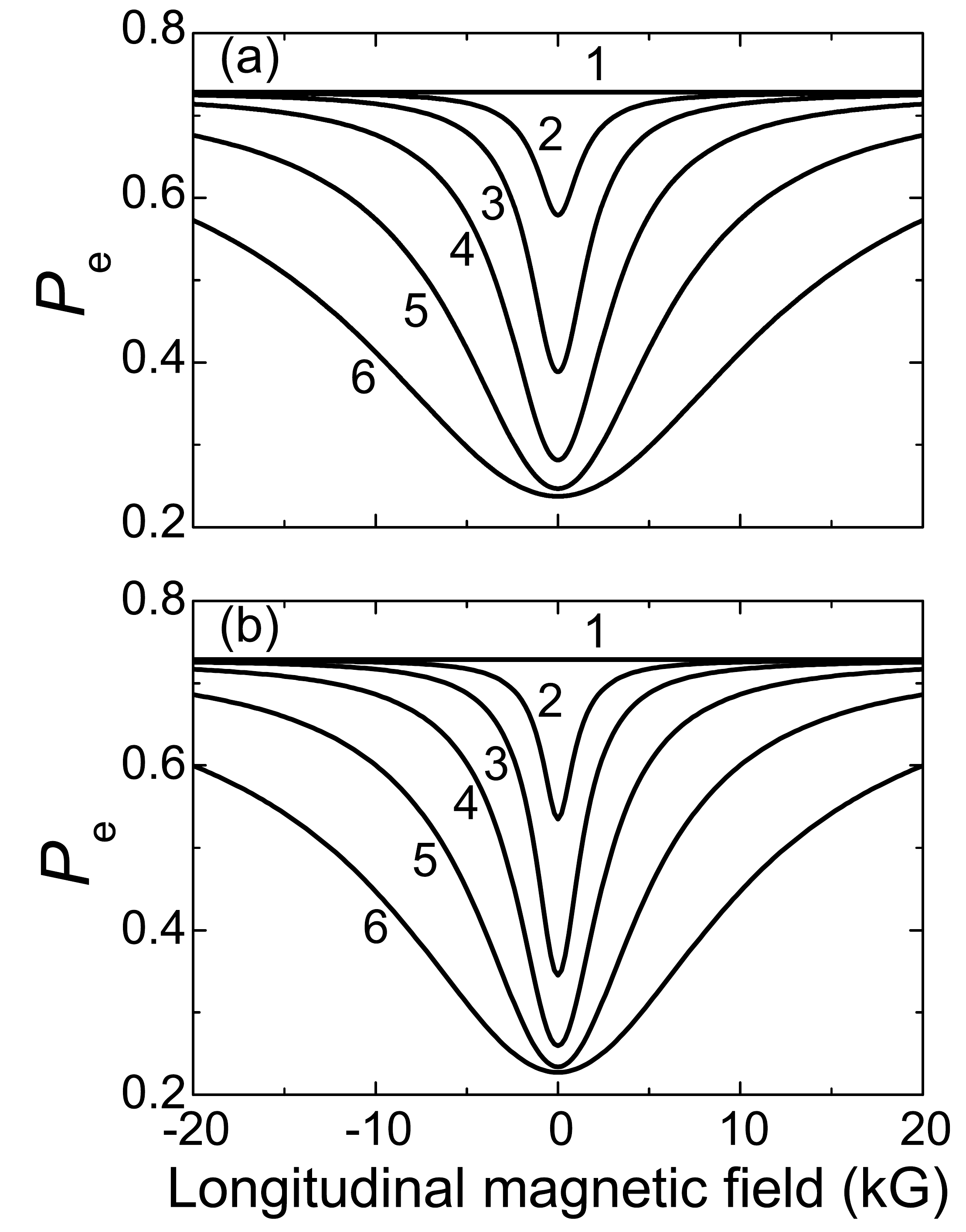}
\caption{\label{fig5} Effect of the hyperfine interaction on the
dependence of the spin polarization of the conduction-band
electrons on the magnetic field. Curves 1 -- 6 are calculated for
the excitation power 60\,mW and the following values of the
hyperfine constant: $A$ = 0 (horizontal line 1), $A_0/4$ (curve
2), $A_0/2$ (3), $A_0$ (4), $2A_0$ (5) and $4A_0$ (6), where
$A_0=17$ $\mu$eV. Panel (a): $\tau_n^{(1)}= \tau_n^{(2)}$ =
150\,ps; panel (b): $\tau_n^{(1)} = 1000$\,ps and $\tau_n^{(2)} =
1$\,ps. The values of other parameters are the same as in Fig.~1.}
\end{figure}

\section{Conclusion}
Thus, due the axial symmetry of the system in the external
longitudinal magnetic field, the components $\rho_{sm,s'm'}$ of
the spin-density matrix of the defect state with a single electron
are nonzero only for the equal total angular momentum projections
$M=s+m$ and $M'=s'+m'$, and the spin-density matrix of the defect
pair-singlet state is diagonal and described by the densities
$N_{2,m}$ of the centers with the nuclear spin projection $m$. The
off-diagonal components $\rho_{s, M - s;s', M - s'}$ with $s \neq
s'$ can be readily expressed via the diagonal components which has
allowed us to derive the quantum master equations containing only
the diagonal components $\rho_{sm,sm}$ and $N_{2,m}$. The
equations take into account the Zeeman splitting of the electron
states in the longitudinal magnetic field, the electron-nuclear
spin coupling described by the hyperfine constant $A$, the spin
relaxation of free and bound electrons described respectively by
the times $\tau_s$ and $\tau_{sc}$, and the nuclear spin
relaxation in the defect states with one and two electrons,
respectively the times $\tau^{(1)}_n$ and $\tau^{(2)}_n$. The
model reproduces the magnetic-field-induced suppression of the
hyperfine interaction, the recovery of the electron spin
polarization and the increase in the edge PL intensity under the
circularly polarized optical excitation. It has been shown that
for the nuclear spin $I = 1/2$ both the PL intensity and circular
polarization are even functions of the longitudinal magnetic field
$B_z$. Moreover, even for $I > 1/2$, there is no shift of
polarization-field or intensity-field curve if the nuclear spin
relaxation is negligible or too fast.

For $I=1/2$ we have calculated the magnetic-field and
excitation-power dependencies of the electron and nuclear spin
polarizations and analyzed the role of the nuclear spin relaxation
in each of the two defect states. The dynamic nuclear spin
polarization appears to be a nonmonotonic function of the
excitation power. Similarly, the ratios $P_e(\infty)/P_e(0)$ and
$J(\infty)/J(0)$ of polarizations and PL intensities at infinitely
high and zero magnetic fields have, as functions of the excitation
power, bell-shaped forms with maxima shifted by several tens of mW
with respect to each other.

\acknowledgments{This research was supported by the RFBR (grant
14-02-00959) and by the Government of Russia through the program
P220 (project 14.Z50.31.0021). We are grateful to K.V. Kavokin,
M.Yu. Petrov, A.Yu. Shiryaev, M.M. Afanasiev and L.S. Vlasenko for
helpful discussions.}

\end{document}